\documentclass[preprintnumbers, prd, twocolumn, showpacs, floatfix, 
preprintnumbers, letterpaper, superscriptaddress,nofootinbib]{revtex4}
\usepackage{graphicx,epsfig}
\usepackage{amsmath}
\usepackage{amssymb}
\usepackage{relsize}
\usepackage{relsize}
\usepackage{relsize}
\usepackage{relsize}

\newcommand{\be}{\begin{equation}}
\newcommand{\ee}{\end{equation}}
\newcommand{\bea}{\begin{eqnarray}}
\newcommand{\eea}{\end{eqnarray}}

\begin{document}

\title{Dynamical features of scalar-torsion theories}

\author{Maria A. Skugoreva}
\email{masha-sk@mail.ru}
\affiliation{Peoples Frendship University of Russia, Moscow 117198,
Russia}

\author{Emmanuel N. Saridakis}
\email{Emmanuel\_Saridakis@baylor.edu}
\affiliation{Physics Division, National Technical University of Athens,
15780 Zografou Campus,  Athens, Greece}
 \affiliation{Instituto de F\'{\i}sica, Pontificia
Universidad de Cat\'olica de Valpara\'{\i}so, Casilla 4950,
Valpara\'{\i}so, Chile}

\author{Alexey V. Toporensky}
\affiliation{Sternberg Astronomical Institute, Lomonosov Moscow State University, Moscow 119992,
Russia}
\affiliation{ Kazan Federal University, Kremlevskaya 18, Kazan 420008, Russia}
\email{atopor@rambler.ru}

\begin{abstract}
We investigate the cosmological dynamics in teleparallel gravity with nonminimal coupling. We analytically extract 
several asymptotic solutions
and we numerically study the exact phase-space behavior. Comparing the obtained results with the corresponding behavior of 
nonminimal scalar-curvature theory, we find significant differences, such is the rare stability and the frequent presence
 of oscillatory behavior.
\end{abstract}


\maketitle

\section{Introduction}

Gravitational modification is one of the main directions one can follow in
order to describe the late-time universe acceleration and/or the early-time
inflationary stage (see for instance  \cite{Nojiri:2006ri,Capozziello:2011et}), which has the
potential advantage of avoiding introducing exotic fields and the concept of
dark energy (see \cite{Copeland:2006wr,Cai:2009zp} and references  therein). 
Definitely one should note that, apart from the different physical
interpretation, one can transform from one approach to the other, partially
or completely, keeping track only of the number of extra degrees of
freedom \cite{Sahni:2006pa}. Thus, one can have various combined scenarios,
with nonminimal couplings between gravity and scalar fields being the most
used class.

Speaking of modified gravity, a natural question arises, namely what
formulation of gravity to use as a basis of modification. The usual approach
on the literature is to start from the standard curvature gravitational
formulation, that is from the Einstein-Hilbert action of General Relativity,
and extend it in various ways  \cite{Nojiri:2006ri,Capozziello:2011et}.
However, a different but still very interesting class of modified gravity
could arise starting from torsional formulations of General Relativity (GR). In
particular, it is well known that Einstein also constructed the
``Teleparallel Equivalent of General Relativity'' (TEGR), where gravity is
described not by the curvature tensor but by the torsion one
\cite{Unzicker:2005in,TEGR,Hayashi:1979qx,JGPereira,Maluf:2013gaa}. The
Lagrangian of this theory is given by contractions of this torsion tensor,
namely the torsion scalar $T$, in a similar way that the Lagrangian of
General Relativity is given by the Ricci scalar $R$, that is from
contractions of the curvature tensor. Hence, instead of modifying GR one
could try to modify TEGR. The most interesting feature is that although GR
coincides with TEGR at the level of equations, their modifications do not,
and thus they correspond to different classes of gravitational modification.

The simplest modification of TEGR is to replace $T$ with $f(T)$ in the
action, resulting to $f(T)$ gravity \cite{Ferraro:2006jd,Linder:2010py}, in a
similar way with the $f(R)$ modification of GR. Since $f(T)$ gravity has no
known curvature equivalent and is a novel modified class, its cosmological 
\cite{Linder:2010py,Chen:2010va,Ong:2013qja} and black hole 
\cite{Bengochea001} applications have attracted significant interest. A next
extension of TEGR arises if one construct and use higher-order torsion
invariants, in a similar way to the use of higher-order curvature invariants
in GR modifications. Thus, constructing the teleparallel equivalent $T_G$ of
the Gauss-Bonnet term $G$, one can build  the $f(T,T_G)$ paradigm
\cite{Kofinas:2014owa}, which is not spanned by the $f(R,G)$ class and thus
is a novel gravitational modification. Furthermore, one could extend TEGR to
$f\left(T,\mathcal{L}_m\right)$ scenario \cite{Harko:2014sja}, with
$\mathcal{L}_m$   the matter Lagrangian, or to $f(T,\mathcal{T})$ theory
\cite{Harko:2014aja}, with $\cal{T}$   the trace of the energy-momentum
tensor, inspired respectively by the $f\left(R,\mathcal{L}_m\right)$
\cite{Bertolami:2007gv,Harko:2008qz} and $f(R,\mathcal{T})$
\cite{Harko:2011kv,Momeni:2011am} extensions of curvature-based gravity,
where again both these theories are different from their curvature
counterparts.

 One could proceed further, and introduce nonminimal couplings in the
framework of TEGR, in a similar way that one introduces these couplings in
GR \cite{Uzan:1999ch}. Thus, in \cite{Geng:2011aj} we formulated the
scenario of ``teleparallel dark energy'', in which $T$ is coupled to an
extra scalar, and as expected this scalar-torsion theory is different from
scalar-(curvature)tensor model, that is from nonminimal quintessence, with
interesting phenomenology \cite{Xu:2012jf,Otalora:2013dsa,Otalora:2013tba,Geng:2013uga,Gonzalez:2014pwa}.
Since scalar-torsion theories are different than scalar-(curvature)tensor ones, in the present work we
are interested in exploring various features of the cosmological dynamics of both theories, and
investigate their differences and similarities. Indeed, we do find that the behavior of the phase space is 
different in the two constructions, and amongst others we find that the scalar-torsion can be free from
run-away solutions, which are typical for the standard curvature-based theory of non-minimally coupled scalar fields.

The plan of the manuscript is outlined as follows: In Section
\ref{model} we briefly present teleparallel equivalent of general relativity and its scalar-torsion extension. In Section \ref{dynanal} we
perform a detailed dynamical analysis, while in Section \ref{comparison} we compare the obtained results with the corresponding behavior of 
nonminimal scalar-curvature theory.
Finally, Section
\ref{Conclusions} is devoted to the conclusions.

\section{Teleparallel and scalar-torsion theory}
\label{model}

In this section  we briefly review the  teleparallel formulation of
General Relativity and its modifications. In this construction the dynamical variables are the
 vierbein  fields ${\mathbf{e}_A(x^\mu)}$, which can be expressed in a
coordinate basis as $\mathbf{e}_A=e^\mu_A\partial_\mu$ \footnote{
Greek indices run over coordinate space-time, while capital Latin indices
span the tangent space-time.}. The vierbeins at each space-time point form an
orthonormal basis for the tangent space, and they are related to the metric
tensor through \begin{equation}  
\label{metrvier}
g_{\mu\nu}=\eta_{AB}\, e^A_\mu \, e^B_\nu,
\end{equation}
where $\eta_{AB}={\rm diag} (1,-1,-1,-1)$. Concerning the independent object
that defines the parallel transportation, i.e. the connection, we use the  
Weitzenb\"{o}ck one $\overset{\mathbf{w}}{\Gamma}^\lambda_{\nu\mu}\equiv
e^\lambda_A\:
\partial_\mu e^A_\nu$ \cite{Weitzenb23}, which leads to zero curvature.
Thus, the gravitational field is described by the torsion tensor, which
reads
\begin{equation}
\label{telelag}
{T}^\lambda_{\:\mu\nu}=\overset{\mathbf{w}}{\Gamma}^\lambda_{
\nu\mu}-%
\overset{\mathbf{w}}{\Gamma}^\lambda_{\mu\nu}
=e^\lambda_A\:(\partial_\mu
e^A_\nu-\partial_\nu e^A_\mu).
\end{equation}
In particular,  constructing the torsion scalar $T$ as 
\begin{equation}
\label{torsionscalar}
T\equiv\frac{1}{4}
T^{\rho \mu \nu}
T_{\rho \mu \nu}
+\frac{1}{2}T^{\rho \mu \nu }T_{\nu \mu\rho }
-T_{\rho \mu }^{\ \ \rho }T_{\
\ \ \nu }^{\nu \mu },
\end{equation}
and using it as a Lagrangian,
variation with respect to the vierbeins leads to exactly the same equations
with General Relativity
\cite{Unzicker:2005in,TEGR,Hayashi:1979qx,JGPereira,Maluf:2013gaa}. That is
why Einstein named this theory ``Teleparallel Equivalent of General
Relativity'' (TEGR).

One can be based on TEGR and start constructing various extensions. As we
discussed in the Introduction, the simplest extension is to replace $T$ by 
$T+f(T)$ in the Lagrangian, i.e. formulating $f(T)$ gravity. Introducing
also the matter sector, the total action will be 
\begin{eqnarray}
\label{actionmatter}
S= \frac{1}{2\kappa^2 }\int d^4x e
\left[T+f(T)\right]+S_m,
\end{eqnarray}
where $S_m$ is the matter action,  $e =
\text{det}(e_{\mu}^A) = \sqrt{-g}$ and $\kappa^2$ is the gravitational constant (we set
the light speed to one). Variation in terms of the vierbeins leads to 
\begin{eqnarray}\label{eom}
&&e^{-1}\partial_{\mu}(ee_A^{\rho}S_{\rho}{}^{\mu\nu})[1+f_{T}]
 +
e_A^{\rho}S_{\rho}{}^{\mu\nu}\partial_{\mu}({T})f_{TT}\ \ \ \ \  \ \ \ \  \ \
\ \ \nonumber\\
&& \ \ \ \
-[1+f_{T}]e_{A}^{\lambda}T^{\rho}{}_{\mu\lambda}S_{\rho}{}^{\nu\mu}+\frac{1}{
4} e_ { A
} ^ {
\nu
}[T+f({T})] \nonumber \\
&&=\frac{\kappa^2}{2} \,e_{A}^{\rho}\overset {em}T_{\rho}{}^{\nu},
\end{eqnarray}
where $f_{T}=\partial f/\partial T$, $f_{TT}=\partial^{2} f/\partial T^{2}$,
 $\overset{em}{T}_{\rho}{}^{\nu}$  denotes the matter
energy-momentum tensor, and we have defined the convenient tensor $
S_\rho^{\:\:\:\mu\nu}\equiv\frac{1}{2}\Big(K^{\mu\nu}_{\:\:\:\:\rho}
+\delta^\mu_\rho
\:T^{\alpha\nu}_{\:\:\:\:\alpha}-\delta^\nu_\rho\:
T^{\alpha\mu}_{\:\:\:\:\alpha}\Big)$, constructed with the help of the 
  contorsion
tensor $K^{\mu\nu}_{\:\:\:\:\rho}\equiv-\frac{1}{2}\Big(T^{\mu\nu}_{
\:\:\:\:\rho}
-T^{\nu\mu}_{\:\:\:\:\rho}-T_{\rho}^{\:\:\:\:\mu\nu}\Big)$. We stress that
although for $f(T)=\text{const}.$ one recovers TEGR with a cosmological constant, and thus GR with a cosmological constant,
for $f(T)\neq\text{const}.$ the
theory is different from $f(R)$ gravity. Thus, $f(T)$ gravity is a new
gravitational modification and that is why it has attracted a significant
interest in the literature
\cite{Linder:2010py,Chen:2010va,Ong:2013qja,Bengochea001}.

An alternative extension of TEGR is to introduce a scalar field
nonminimally coupled with $T$, in a similar way that in curvature-based
gravity one introduces a scalar field nonminimally coupled to $R$. In
particular, the total action will be  
\cite{Geng:2011aj} 
\begin{equation}
\!\! S=\int
d^{4}x\,e\,\left[\frac{T}{2\,\kappa^2}+\frac{1}{2}\,\partial_{\mu}\phi\,
\partial^{\mu}\phi-V(\phi)+\frac{\xi}{2}\,B(\phi)\,T+\mathcal{L}_m\right] ,
 \label{totalaction}
\end{equation}
where $\phi$ is a canonical scalar field, $V(\phi)$ its potential, and
$B(\phi)$ its arbitrary nonminimal coupling with the torsion scalar $T$. 
Variation with respect to the vierbein yields the coupled field
equation
\begin{eqnarray}
&&\!\!\!\!\!\!\!\!\!\!\! \left[\frac{2}{\kappa^{2}}+2 \xi\,B(\phi)\right]\left[
e^{-1}\partial_{\mu}(ee_A^{\rho}S_{\rho}{}^{\mu\nu})
-e_{A}^{\lambda}T^{\rho}{}_{\mu\lambda}S_{\rho}{}^{\nu\mu}+\frac{1}{
4} e_{A}^{\nu}T
\right]\nonumber
\\
&&\ \ \ \ \ \ \ \ \ \  \,
-
e_{A}^{\nu}\left[\frac{1}{2}
\partial_\mu\phi\partial^\mu\phi-V(\phi)\right]+
  e_A^\mu \partial^\nu\phi\partial_\mu\phi,
\nonumber
\\
&&\ \ \ \ \ \ \ \ \ \  \, + 2\xi
e_A^{\rho}S_{\rho}{}^{\mu\nu}B'(\phi)
\left(\partial_\mu\phi\right)= e_{A}^{\rho}\overset
{\mathbf{em}}T_{\rho}{}^{\nu},
 \label{eoms}
\end{eqnarray}
where primes denote the derivative of a function with respect to its argument.

In order to focus on the cosmological application of this theory, we impose 
the vierbein ansatz \begin{equation}
\label{veirbFRW}
e_{\mu}^A={\rm
diag}(1,a(t),a(t),a(t)),
\end{equation}
which leads to the flat Friedmann-Robertson-Walker (FRW)
metric
\begin{equation}
ds^2= dt^2-a^2(t)\,\delta_{ij} dx^i dx^j \,,
\end{equation}
where $a(t)$ is the scale factor.
Thus, with this vierbein ansantz, the equations of motion (\ref{eoms})
give rise to the modified Friedmann equations
\begin{equation}
\label{Fr1}
3H^2=\kappa^2\left[\frac{{\dot{\phi}}^2}{2}+V(\phi)-3\xi H^2 B(\phi)+\rho_m\right],
\end{equation}
\begin{equation}
\label{Fr2}
2\dot{H}=-\kappa^2\left[{\dot{\phi}}^2+2\xi H B'(\phi)
\dot{\phi}+2\xi\dot{H}B(\phi)+\rho_m(1+\omega_m)\right],
\end{equation}
where $H=\dot{a}/a$ is the Hubble function, and dots denote differentiation with
respect to $t$. Note that in the above relations we have   used  the
relation $T=-6H^2$, which holds for the vierbein choice
(\ref{veirbFRW}). Additionally, we have considered the matter Lagrangian to
correspond to a perfect fluid with energy density and pressure $
\rho_m$ and $p_m$  respectively, and we have defined its equation-of-state
parameter to be $w_m\equiv p_m/\rho_m$. Finally, varying the action 
(\ref{totalaction}) with respect to $\phi$, and imposing the FRW ansatz, we
acquire the scalar field equation of motion
\begin{equation}
\label{phieom}
\ddot{\phi}+3 H\dot{\phi}+3\xi H^2 B'(\phi)+V'(\phi)=0.
\end{equation}
Lastly, note that, as expected, for $\xi=0$ the scenario at hand coincides 
completely with standard quintessence, since in this case the scalar field is minimally coupled to TEGR, which in turn 
coincides with General Relativity.

 \section{Dynamical Analysis}
 \label{dynanal}
 
In the cosmological application of any gravitational theory
one can find many analytical solutions, and an infinite number of 
numerical ones. However, the most important issue is to investigate the global features of the dynamics, that is
to extract information about the given cosmological model that is independent of the initial conditions 
and the specific evolution. This is obtained using the powerful method of dynamical analysis, which allows to 
examine in a systematic way all the possible asymptotic cosmological behaviours, that is all the possibilities
of the universe behaviour at late times. In particular, if stable late-time solutions are revealed, it is implied that
the universe will result to them independently of the initial conditions and the model parameters.

 The phase-space and stability analysis is performed by transforming the given cosmological model
 into its autonomous form, which in general will be of the form $d\textbf{Y}/d\ln a=\textbf{f(Y)}$, 
where $\textbf{Y}$ is a vector constituted by suitable chosen 
variables and $\textbf{f(Y)}$ the corresponding   
vector of the autonomous equations \cite{Copeland:1997et,Leon2011}. Then the critical points $\bf{Y_c}$  of this dynamical system 
are extracted imposing the condition $d\textbf{Y}/d\ln a=0$. Thus, in order to examine
the stability of these critical points one expands the system around
$\bf{Y_c}$ as
$\bf{Y}=\bf{Y_c}+\bf{U}$, with $\textbf{U}$ the column vector of the variable perturbations, and for 
each
critical point he expands the equations for the perturbations up to
  first order as $  \textbf{U}'={\bf{Q}}\cdot
\textbf{U}$, where the matrix ${\bf {Q}}$ contains all the coefficients of the
perturbation equations. Hence, the stability properties and the type of a specific critical point are determined by the 
 eigenvalues
of ${\bf {Q}}$, namely  if all eigenvalues   have positive
real parts  then this point will be unstable, if they all have negative
real parts then it will be stable, while if they change sign   it will be a
saddle point.

Let us apply the above method to the scalar-torsion cosmology that was presented in the previous section. As an example
we will focus on power-law potentials and power-law coupling functions, since the exponential cases have been investigated 
elsewhere \cite{Xu:2012jf,Otalora:2013tba}. 
In particular, in the following we consider the case where 
\begin{equation}
\label{potpowerlaw}
V(\phi)=V_0 \phi^n
\end{equation}
and 
\begin{equation}
\label{Bpowerlaw}
B(\phi)=\phi^N,
\end{equation}
where we focus on even $n$   in order for the potential to be non-negative. For convenience we choose $N>0$, although the 
incorporation of the $N<0$ case is straightforward. 
 Additionally, since for $\xi=0$ the scenario at hand coincides completely with 
 standard quintessence, in the following we focus on the case of interest of this work, that is on $\xi\neq0$. 
The dynamical analysis proves to be different for the cases $N\neq2$ and $N=2$, and thus in the following subsections we examine these
cases separately.


\subsection{$N\neq2$} 

In order to proceed, and as we described above, we introduce the following dimensionless auxiliary variables:
\begin{eqnarray}
\label{auxiliary1}
&&x=\frac{\kappa^2{\dot{\phi}}^2}{6 H^2[1+\kappa^2\xi B(\phi)]}\\
\label{auxiliary2}
&&
y=\frac{\kappa^2 V(\phi)}{3 H^2[1+\kappa^2\xi B(\phi)]}\\
\label{auxiliary3}
&&
z=\frac{\kappa^2 \rho_m}{3 H^2[1+\kappa^2\xi B(\phi)]}\\
\label{auxiliary4}
&&
m=\frac{\dot{\phi}}{H \phi}\\
\label{auxiliary5}
&&A=\frac{B'(\phi)\phi}{1+\kappa^2\xi B(\phi)}=\frac{N}{1/\phi^N+\kappa^2\xi}.
\end{eqnarray}
Note the useful relation between $A$, $x$, $m$, namely
\begin{equation}
\label{Axmrelation}
6 N x{(N-\kappa^2\xi A)}^{\frac{2-N}{N}}=\kappa^2 A^{\frac{2}{N}}m^2.
\end{equation}     
Furthermore, it proves convenient to additionally define the following functions, where in our case become just 
 dimensionless parameters:
\begin{eqnarray}
&&b=\frac{B''(\phi)\phi}{B'(\phi)}=N-1\label{bdef}\\
&&
c=\frac{V'(\phi)\phi}{V(\phi)}=n.
\end{eqnarray}
In terms of the  auxiliary variables the first Friedmann equation (\ref{Fr1})
gives rise to the following constraint 
\begin{equation}
\label{constraint}
1=x+y+z,
\end{equation}
while using additionally the above parameters and the scalar-field evolution equation 
(\ref{phieom}), we can obtain the following expressions for the physically interesting quantities:
\begin{eqnarray}
\label{phidotdot}
&& \frac{\ddot{\phi}}{H\dot{\phi}}=-3-\frac{m}{2 x}(\kappa^2\xi A+n y)\\
&&\frac{\dot{H}}{H^2}=-3 x-\kappa^2\xi A m.
\label{HdotH2}
\end{eqnarray}

Since we are interested in investigating the dynamical features of the pure modified gravitational sector, 
we focus on the vacuum case $\rho_m=0$. Hence $z=0$ and the dimensionality of the phase space
is reduced by one. In this case, the cosmological equations of the scenario at hand  are finally transformed 
 to their autonomous form     
\begin{eqnarray}
\label{eq1}
&&\frac{dm}{d\ln a}=
-\frac{3 N(\kappa^2\xi A+n){(N-\kappa^2\xi A)}^{\frac{2-N}{N}}}{\kappa^2
A^{\frac{2}{N}}}\nonumber\\
&&
\ \ \ \ \ \ \ \, \ \ \ \ +\frac{\kappa^2
A^{\frac{2}{N}}m^3}{2N{(N-\kappa^2\xi A)}^{\frac{2-N}{N}}}\nonumber\\
&&
\ \ \ \ \ \ \ \ \ \ \ \,
+m^2\left(\frac{n}{2}+\kappa^2\xi A
-1\right)-3m
\\
&&\frac{dA}{d\ln a}=A m(N-\kappa^2\xi A),
\label{eq2}
\end{eqnarray} 
where we have replaced $b+1=N$ as it arises from (\ref{bdef}).

The system (\ref{eq1}),(\ref{eq2}) admits only one real critical point:

\subsubsection{ \textbf{Point $P_1$:  $m=0$, $A=-\frac{n}{\kappa^2\xi}$}}

The   critical point $P_1$ corresponds to  $m=0$, $A= -\frac{n}{\kappa^2\xi}$, 
which according to (\ref{Axmrelation}) leads to $x=0$, and then using (\ref{constraint}) to $y=1$.
 Since $A=\text{const}.$ from (\ref{auxiliary5}) we deduce that the scalar field is also constant at $ P_1$, namely
\begin{equation}
 {\phi}^N={\phi_0}^N=-\frac{n}{\kappa^2\xi(N+n)},
\end{equation}
 while according to  (\ref{HdotH2}) we find that $\dot{H}/H^2$ is  zero. Thus, knowing additionally that  $y$ is non-zero,  from
(\ref{auxiliary2}) we deduce that the Hubble function obtains a constant non-zero value at $P_1$, namely
 \begin{equation}
 H=H_0=\sqrt{-\frac{nV_0{\phi_0}^{n-N}}{3 N\xi}},
\end{equation} 
and thus the universe is in a de Sitter phase where the scale factor 
expands exponentially as
\begin{equation}
a(t)=a_0{e}^{H_0(t-t_0)},
\end{equation}
where   $a_0$,$t_0$ are constants.

Linearizing the perturbation equations around $P_1$ and extracting the eigenvalues of the corresponding 
perturbation matrix, as we discussed earlier, we find:
{\small{
\begin{eqnarray}
&&\!\!\!\!\!\!\!\!\!\!\!\!\!\!\!\!\!\!\!\lambda_{1,2}=-\frac{3}{2}
 \pm\frac{\sqrt{3}}{2\kappa^2}{\left(-\frac{n}{\kappa^2\xi}\right)}^{-\frac{N}{2}}\nonumber\\
 &&\!\!\!\!\!\!\!\!\!\!\!\!\!\!\!\!\!\!\!\cdot
\left\{\frac{n\kappa^2}{
\xi}\left[4\xi N{\left(-\frac{n}{\kappa^2\xi}\right)}^{\frac{N}{2}}{\left(N+n\right)}^{\frac{2}{N} }
-3{\left(-\frac{n}{\kappa^2\xi}\right)}^{N-1}
\right]\right\}^{\frac{1}{2}}\!.
\label{eigenP1}
\end{eqnarray} 
}}
Due to the complexity of the above expressions it is not possible to derive analytical results for the signs of
the real parts of these eigenvalues. Thus, we perform a numerical scanning in the parameter space.
\begin{figure*}
\text{~~~~}\includegraphics [scale=0.88] {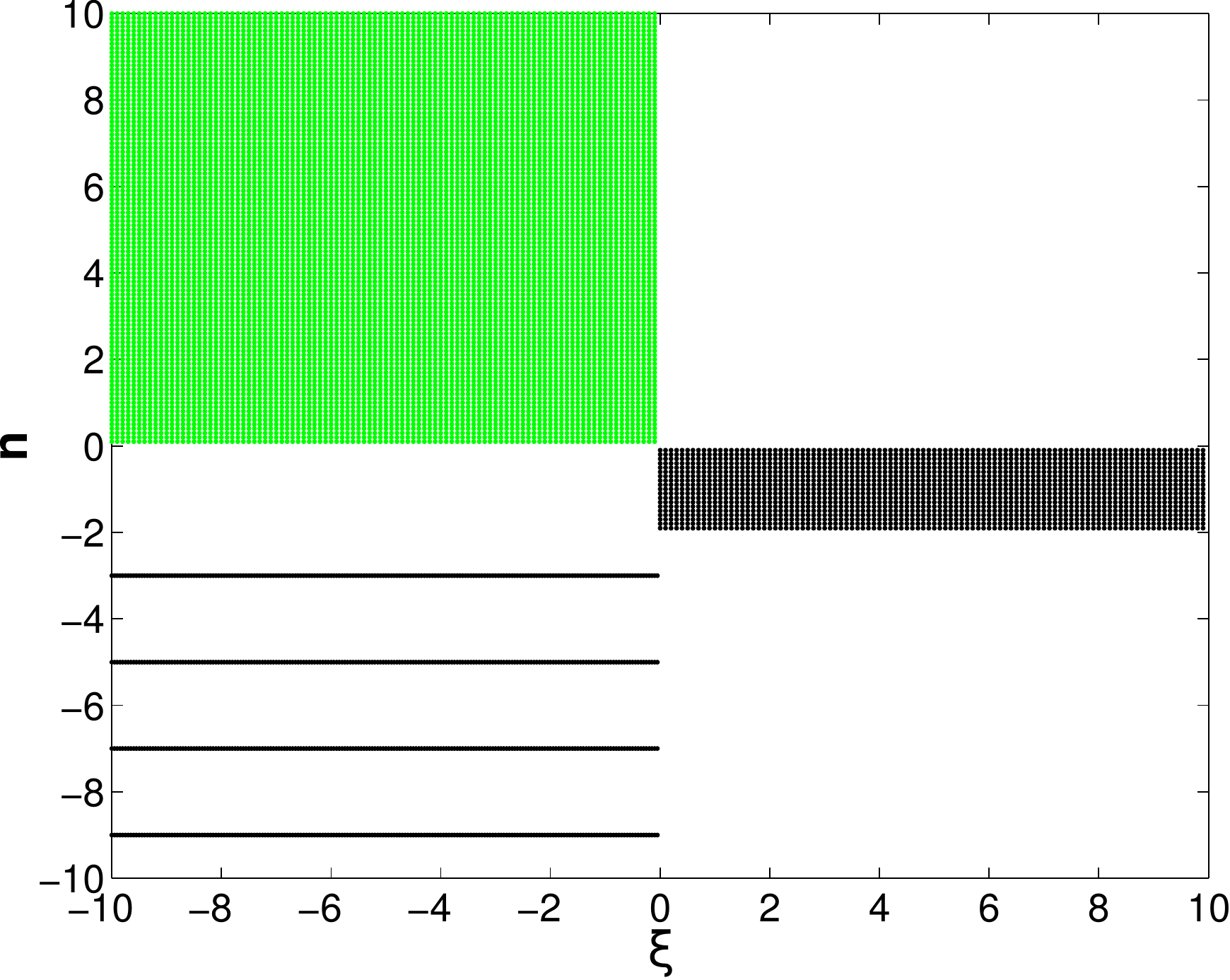}
\caption{
{\it{
Stability regions of de Sitter point $P_1$ in the parameter subspace $(n,\xi)$ for $N=2$. Green regions correspond to 
saddle behavior, while black regions  denote stable behavior. }}   }
\label{N2}
\end{figure*}
\begin{figure*}
\text{~~~~} 
\includegraphics [scale=0.88] {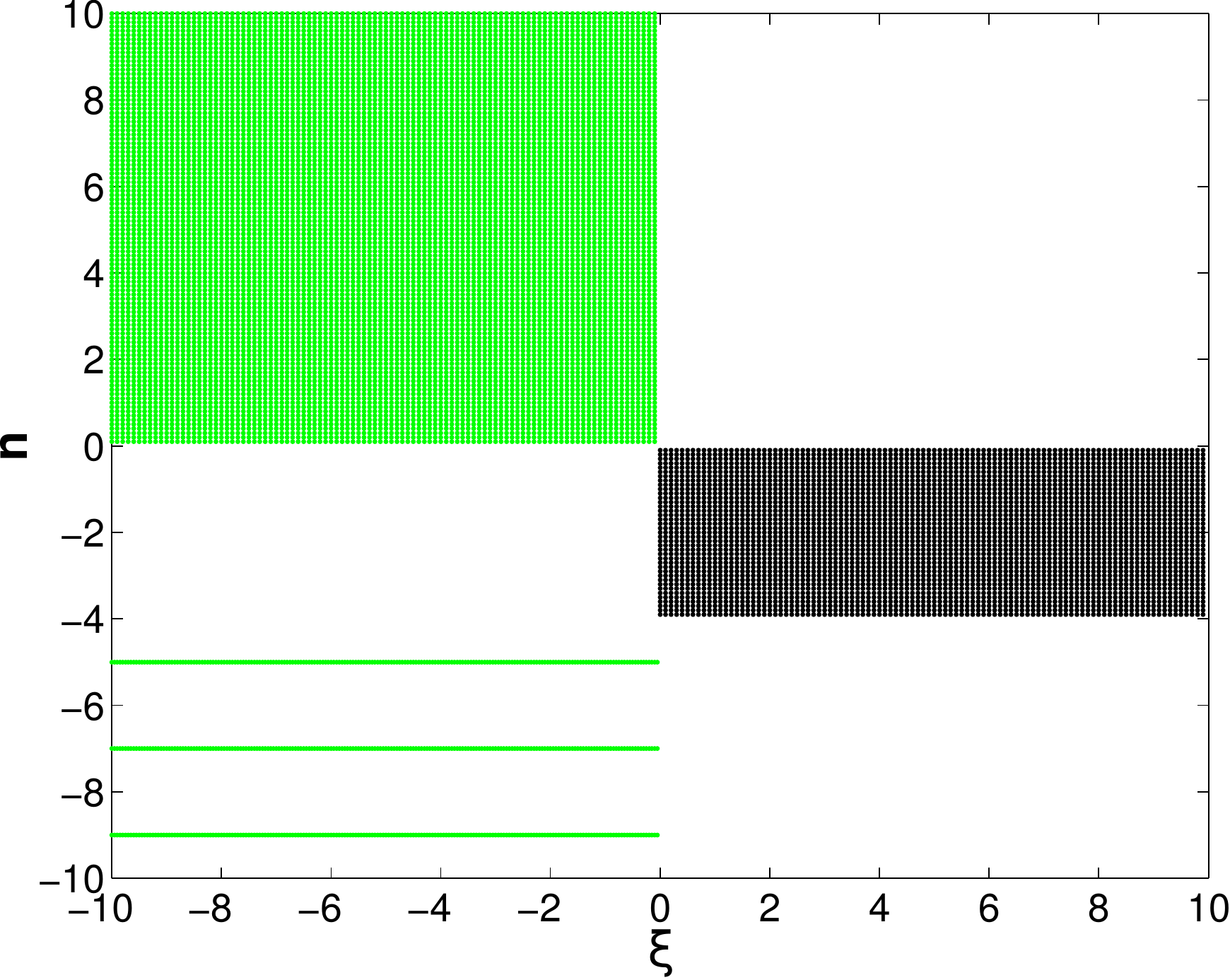}
\caption{
{\it{
Stability regions of de Sitter point $P_1$ in the parameter subspace $(n,\xi)$ for $N=4$. Green regions correspond to 
saddle behavior, while black regions  denote stable behavior. }}   }
\label{N4}
\end{figure*}
\\

In Figs. \ref{N2} and \ref{N4} we  present the stability regions in the parameter subspace  $(n,\xi)$ for    $N=2$ and  $N=4$
respectively. 
Since for $\xi<0$, $n<0$ the de Sitter solution exists only for negative $\phi_0^{n-N}$ (see Eq. (29)), we restrict ourselves by
odd integer $|n|$, in other quadrants $n$ can be a continious variable (in white zones the de Sitter solution does not exist). 
As we observe, in the case where both $N$ and $n$
are positive
we find that there
is not any stability region, and thus point $P_1$ is saddle. However, in the case where $n<0$, and for $N>2$, we do find stability regions,
that is in this case the de Sitter point $P_1$ is stable. We do not consider negative $n$ further in the present paper.

\subsection{$N=2$} 
\label{N22}

In this case we also use the auxiliary variables (\ref{auxiliary1})-(\ref{auxiliary5}), the constraint (\ref{constraint})
remains the same, but   now    relation 
(\ref{Axmrelation}) simplifies to 
\begin{equation}
\label{Axmrelationb}
x=\frac{\kappa^2 Am^2}{12}.
\end{equation}
Moreover, instead of (\ref{phidotdot}),(\ref{HdotH2}) we now have
\begin{eqnarray}
 \label{phidotdotb}
 &&\frac{\ddot\phi}{H\dot\phi}=-3-6\frac{\xi}{m}-6\frac{n}{\kappa^2 m A}+\frac{nm}{2}\\
&&\frac{\dot H}{H^2}=-\frac{\kappa^2 A m^2}{4}-\kappa^2\xi A m.
 \label{HdotH2b}
\end{eqnarray}
Hence, the autonomous form of the cosmological system   becomes
\begin{eqnarray}
\label{eq1b}
&&\frac{dm}{d\ln a}=
-3 m-6\xi-\frac{6n}{\kappa^2 A}+ \frac{\kappa^2 A m^3}{4} \nonumber\\
&&\ \ \ \ \ \ \ \ \ \ \ \, +m^2\left(\frac{n}{2}+\kappa^2\xi A
-1\right)
\\
\label{eq2b}
&&\frac{dA}{d\ln a}=Am(2-\kappa^2\xi A).
\end{eqnarray}

The system (\ref{eq1b}),(\ref{eq2b}) admits four real critical points:

\subsubsection{ \textbf{Point $Q_1$:  $m=0$, $A=-\frac{n}{\kappa^2\xi}$}}

The first critical point $Q_1$ corresponds to  $m=0$, $A=-\frac{n}{\kappa^2\xi}$, 
which according to (\ref{Axmrelationb}) leads to $x=0$, and then using (\ref{constraint}) gives $y=1$ (again we focus on the case
$\rho_m=0$ i.e. $z=0$). Hence, this point can be obtained from point $P_1$ of the previous subsection, setting  $N=2$. In particular,
from the definition of $A$ in 
   (\ref{auxiliary5}) we find that the scalar field is constant too, namely
   \begin{equation}
    {\phi}^2=    {\phi_0}^2= -\frac{n}{\kappa^2\xi(2+n)},
   \end{equation}
while from (\ref{HdotH2b}) we deduce that 
 $\dot{H}/H^2$ is  zero. Thus, knowing additionally that  $y$ is non-zero,  from
(\ref{auxiliary2}) we deduce that the Hubble function obtains a constant non-zero value at $Q_1$, namely
 \begin{equation}
 H=H_0=\sqrt{-\frac{nV_0{\phi_0}^{n-2}}{6\xi}}.
\end{equation} 
Therefore, the universe is in a de Sitter phase where the scale factor 
expands exponentially as
\begin{equation}
a(t)=a_0{e}^{H_0(t-t_0)}.
\end{equation}
 Note that for growing potentials ($n>0$) this de Sitter fixed point   exists   only for $\xi<0$.
 
The eigenvalues of the perturbation matrix can be obtained from  (\ref{eigenP1}) setting $N=2$, and thus we have  
\begin{eqnarray}
 \lambda_{1,2} =-\frac{3}{2}\pm\frac{1}{2}\sqrt{9-24\xi(n+2)}.
\end{eqnarray}
Therefore, we can easily see that this point is always a saddle one.

\subsubsection{ \textbf{Point $Q_2$:  $m=\sqrt{6\xi}$, $A=\frac{2}{\kappa^2\xi}$}}

The second critical point $Q_2$ exists only for $\xi>0$ and corresponds to  $m=\sqrt{6\xi}$, $A=\frac{2}{\kappa^2\xi}$,
and thus according to (\ref{Axmrelationb}) we get $x=1$, and then using (\ref{constraint}) we obtain $y=0$. Therefore, from
 (\ref{HdotH2b}) we find that 
  \begin{equation}
  \label{HdotH2Q2}
\frac{\dot H}{H^2}=-3-2\sqrt{6\xi},
\end{equation} 
which leads to 
\begin{equation}
a(t)=a_0{|t-t_0|}^{\frac{1}{3+2\sqrt{6\xi}}},
\end{equation}
where   $a_0$,$t_0$ are constants. Note that at $t\rightarrow\infty$ both $H$ and $\dot{H}$ tend to zero  
($H\sim t^{-1}$ and $\dot{H}\sim t^{-2}$), but with the ratio $H/\dot{H}^2$ being the constant given in 
(\ref{HdotH2Q2}). Inserting the scale-factor expression into the $m$-definition in (\ref{auxiliary4}), we extract the 
scalar-field solution as
 \begin{equation}
\phi(t)=\phi_0{|t-t_0|}^{\frac{\sqrt{6\xi}}{3+2\sqrt{6\xi}}}.
\end{equation}
Finally, substituting the coordinates of $Q_2$ into (\ref{auxiliary5}) we deduce that $\phi\rightarrow\infty$,
and thus the only case is when $t_0<t\rightarrow\infty$. Hence point $Q_2$ cannot give rise to a Big Rip  \cite{Sami:2003xv}.

Lastly, substituting these expressions into the $y$-definition in (\ref{auxiliary2}), 
we deduce that in order to have self-consistency the potential must have $n<-2$, which is not of interest for the scope of the present work.

\subsubsection{\textbf{Point $Q_3$:  $m=-\sqrt{6\xi}$, $A=\frac{2}{\kappa^2\xi}$}}

The third critical point $Q_3$ exists only for $\xi>0$ and corresponds to $m=-\sqrt{6\xi}$, $A=\frac{2}{\kappa^2\xi}$.
Thus, from (\ref{Axmrelationb}) we acquire $x=1$, and then from (\ref{constraint}) we get $y=0$.
From
 (\ref{HdotH2b}) we find that 
  \begin{equation}
   \label{HdotH2Q3}
\frac{\dot H}{H^2}=-3+2\sqrt{6\xi},
\end{equation} 
which, for $\xi\neq3/8$, leads to 
\begin{equation}
a(t)=a_0{|t-t_0|}^{\frac{1}{3-2\sqrt{6\xi}}},
\end{equation}
where   $a_0$,$t_0$ are constants.  Note that at $t\rightarrow\infty$ both $H$ and $\dot{H}$ tend to zero 
($H\sim t^{-1}$ and $\dot{H}\sim t^{-2}$), but with the ratio $H/\dot{H}^2$ being the constant given in 
(\ref{HdotH2Q3}). Inserting the scale-factor expression into the $m$-definition in (\ref{auxiliary4}), we obtain the 
scalar-field solution as
 \begin{equation}
\phi(t)=\phi_0{|t-t_0|}^{\frac{\sqrt{6\xi}}{2\sqrt{6\xi}-3}}.
\end{equation}
Substituting these expressions into   (\ref{auxiliary2}), and contrary to point $Q_2$, we deduce that self-consistency is obtained 
for both positive and negative values of $n$. Restricting ourselves to the positive-$n$ cases, we extract the following requirements for the existence
of this critical point: 
\begin{eqnarray}
 \text{For}\ n<2 \ \text{it exists for} &&\  \xi\in\left( 0;\frac{3}{8}\right) \cup\left( \frac{3}{8};\frac{6}{{(n+2)}^2}\right) \nonumber\\
  \text{For}\ 2\leq n\ \text{it exists for} &&\ 0<\xi < \frac{6}{(n+2)^2}\nonumber.
\end{eqnarray}
Concerning its stability, the eigenvalues of the perturbation matrix are found to be
\begin{eqnarray}
&& \lambda_1 =2\sqrt{6\xi}\nonumber\\
&& \lambda_2=6-\sqrt{6\xi}(n+2),
\label{eigenQ3}
\end{eqnarray}
and thus $Q_3$ is always an unstable node, since $\sqrt{\xi}(n+2)<\sqrt{6}$.
 
 Finally, in the case $\xi=3/8$ with $n<2$ equation (\ref{HdotH2Q3}) leads to  
  the trivial solution $H=H_0=const$, and thus  for the scale factor we acquire  
  \begin{equation}
  \label{aexpQ3}
 a(t)=a_0 e^{H_0(t-t_0)},
 \end{equation}
while relation (\ref{auxiliary4}) leads to 
  \begin{equation} 
\phi(t)=\phi_0 e^{ -\sqrt{6\xi}H_0(t-t_0)}.
  \label{phiexpQ3}
\end{equation} 
In this case the eigenvalues of the perturbation matrix are still given by (\ref{eigenQ3}) but for $\xi=3/8$, and thus
we deduce that   $Q_3$  is  an unstable node.

\subsubsection{ \textbf{Point $Q_4$: 
$m=-\xi(n+2)$}, $A=\frac{2}{\kappa^2\xi}$}

The fourth critical point corresponds to $m=-\xi(n+2)$, $A=\frac{2}{\kappa^2\xi}$, and   
 thus according to (\ref{Axmrelationb}) we get $x=\frac{\xi{(n+2)}^2}{6}$, and then using (\ref{constraint}) we 
 acquire $y=1-\frac{\xi{(n+2)}^2}{6}$.
 Hence, from
 (\ref{HdotH2b}) we find that 
  \begin{equation}
  \label{HdotQ4}
\frac{\dot H}{H^2}=\frac{-\xi(n+2)(n-2)}{2}.
\end{equation} 

For  $n\neq2$ the above equation leads to 
\begin{equation}
a(t)=a_0{|t-t_0|}^{\frac{2}{\xi(n^2-4)}},
\label{aQ4}
\end{equation}
where   $t_0$ is a  constant. Then, inserting this expression into the $m$-definition in (\ref{auxiliary4}), we extract the 
scalar-field solution as
 \begin{equation}
 \label{phisolQ4}
\phi(t)=\phi_0{|t-t_0|}^{\frac{2}{2-n}}.
\end{equation}
 
As $A=\frac{2}{\kappa^2\xi}$ then $\phi\to\infty$ at this stationary point.

Substituting found solution (51), (52) to the initial system (10), (11), (12) we get corresponding conditions of existence ($V_0<0$ is not considered in this paper): 
\\for $N=2$ and\\
1). $n\neq2$ --- even, $V_0>0$, $0<\xi<\frac{6}{{(n+2)}^2}$, $\forall \phi_0$\\ 
2). $n\neq2$ --- odd, $V_0>0$, $0<\xi<\frac{6}{{(n+2)}^2}$, $\phi_0>0$ or\\
$V_0>0$, $\xi\in(-\infty;0)\cup(\frac{6}{{(n+2)}^2};+\infty)$, $\phi_0<0$.\\


  The eigenvalues of the perturbation matrix for $Q_4$ read  
\begin{eqnarray}
&& \lambda_1 =2\xi(n+2)\nonumber\\
&& \lambda_2=-3+\frac{\xi{(n+2)}^2}{2},
\label{eigenQ4}
\end{eqnarray}
This point for $\xi>\frac{6}{{(n+2)}^2}$ --- an unstable node, ~~for $0<\xi<\frac{6}{{(n+2)}^2}$ --- a saddle, ~~for $\xi<0$ -- a stable node.

  For $n=2$, equation (\ref{HdotQ4}) does not accept the solution (\ref{aQ4}), but it has the trivial solution $H=H_0=const$, which inserted into $y$-definition 
  relation (\ref{auxiliary2}) gives $H_0^2=\frac{V_0}{\xi(3-8\xi)}$. Since we focus on positive potential, i.e. with  $V_0>0$, we deduce that this solution exist 
  for $0<\xi<3/8$. Hence, for the scale factor we obtain  
  \begin{equation}
  \label{aexpQ4}
 a(t)=a_0 e^{\sqrt{\frac{V_0}{\xi(3-8\xi)}}(t-t_0)},
 \end{equation}
while relation (\ref{auxiliary4}) gives 
  \begin{equation} 
\phi(t)=\phi_0 e^{ \sqrt{\frac{16V_0\xi}{(3-8\xi)}}(t-t_0)}.
  \label{phiexpQ4}
\end{equation} 
In this case the eigenvalues of the perturbation matrix are still given by (\ref{eigenQ4}) but for $n=2$, and since $0<\xi<3/8$
we deduce that $Q_4$ for $n=2$ is always a saddle point.

\section{Comparison between scalar-torsion and scalar-curvature behavior}
\label{comparison}

In the previous section we performed the dynamical analysis of
some general scalar-torsion models. In this section we present specific figures
in the phase space, and we discuss the physical features of the obtained cosmology.
Then, we compare these results with the known behavior 
of the corresponding scalar-curvature scenarios \cite{Sami, Vernov}. Since in scalar-curvature models  
only the case of positive $\xi$ has been studied in detail, in the following we focus on this case too. 
 Similar to the scalar-curvature models the important regime of dumped scalar field oscillations can not be extracted 
from fixed points analysis for  the set of variables we used.

As it is usual  in the majority of cosmological scenarios,
the scalar field and the scale factor diverge at the critical points, and therefore for presentation reasons
it proves more convenient to define suitable compact variables, projecting the dynamics in the unit circle. 
In particular, we define
\begin{equation}
\alpha=\frac{\phi}{\sqrt{1+{\phi}^2+{\dot{\phi}}^2}},
~~~~\beta=\frac{\dot{\phi}}{\sqrt{1+{\phi}^2+{\dot{\phi}}^2}},
\end{equation}
and thus the inverse transformation reads
\begin{equation}
\label{alphabeta}
\phi=\frac{\alpha}{\sqrt{1-{\alpha}^2-{\beta}^2}},
\qquad\dot{\phi}=\frac{\beta}{\sqrt{1-{\alpha}^2-{\beta}^2}}.
\end{equation}
Obviously we have 
\begin{equation}
\label{absystem}
\begin{split}
\dot{\alpha}&=\beta\left(1-\alpha^2-\alpha\ddot{\phi}\sqrt{1-{\alpha}^2-{\beta}^2}\right),
\\
\dot{\beta}&=\ddot{\phi}\sqrt{1-{\alpha}^2-{\beta}^2}(1-{\beta}^2)-\alpha{\beta}^2.
\end{split}
\end{equation}

\subsection{$N\neq2$} 

\begin{figure}[hbtp] 
\includegraphics [scale=0.44] {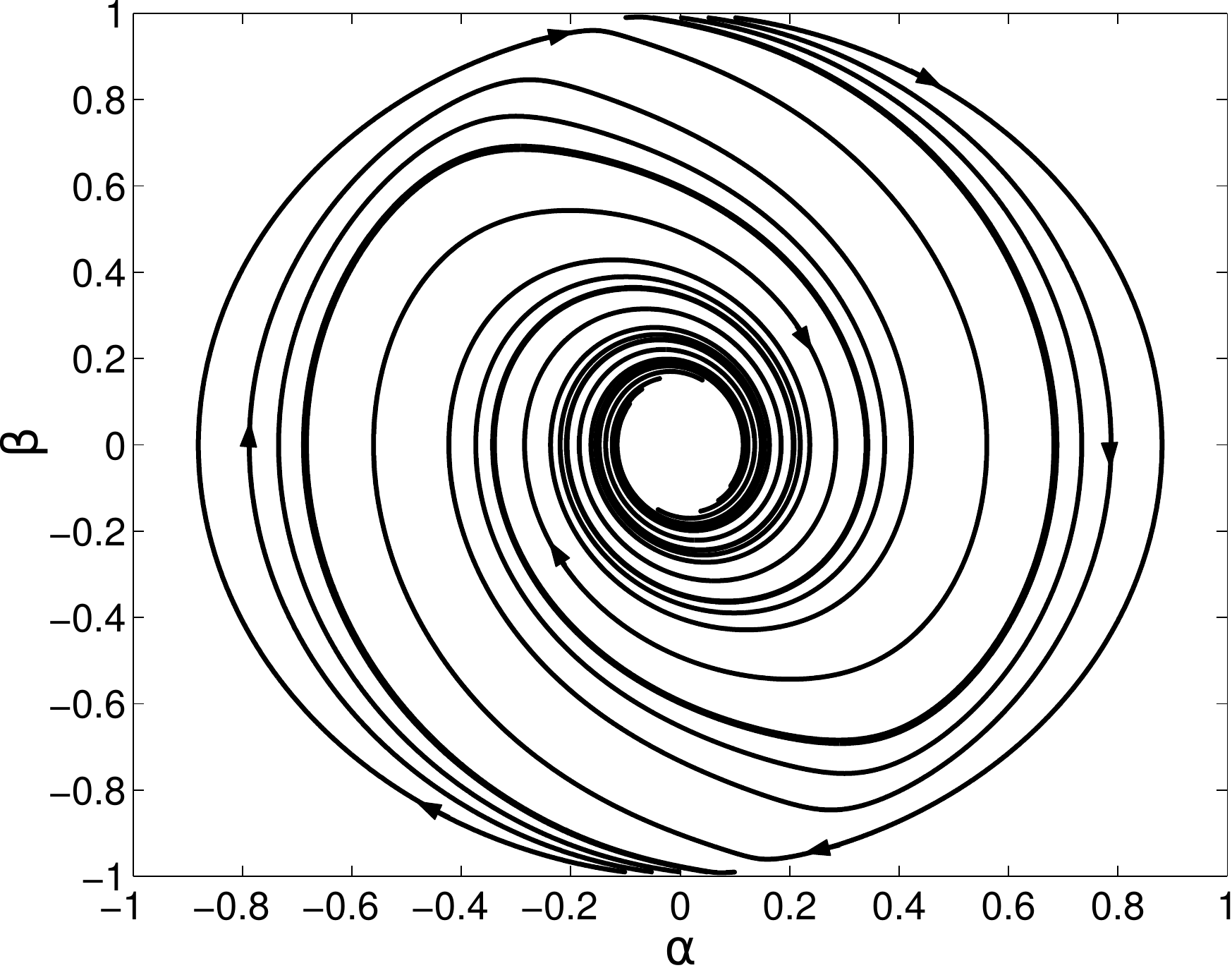}
\caption{{\it{
The projection of the phase-space evolution of the system (\ref{eq1}),(\ref{eq2}) on the  $(\alpha,\beta)$ 
plane, in the case where   $N=4$, $n=2$, $\xi=1$, with $V_0=1$,  $\kappa^2=1$, i.e.
in the case of  quartic nonminimal coupling function   and quadratic potential.
There are no stable fixed points, and the scalar field exhibits oscillations. }} }
\label{Fig5}
\end{figure}

We start by examining the phase portraits for the $N\neq2$ case. As we discussed in the previous section, the only critical point in 
this case is the de Sitter point $P_1$, but since it exists only for $\xi<0$ we are not going to discuss it in detail. Hence, remaining in the case 
$\xi>0$ we conclude that the only possible evolution is that of scalar-field oscillations, and the  global
picture of such dynamics does not depend on the particular value of $\xi$. In order to  present this behavior 
more transparently, in  Fig.~\ref{Fig5} we depict the phase-space evolution in the $(\alpha,\beta)$ plane, in the case where  $N=4$, $n=2$, $\xi=1$, with $V_0=1$, $\kappa^2=1$. 
Note that this behavior is similar to that in scalar-curvature case \cite{Sami, Vernov}.

\subsection{$N=2$} 

We now examine the more interesting $N=2$ case.   
For this quadratic coupling function, the quadratic potential $n=2$ is an exceptional one, since in this case 
the obtained solutions are not power laws but exponentials.
Having in mind the existence and the stability conditions of points $Q_1$ to $Q_4$ of subsection \ref{N22}, we deduce that 
when $n=2$,
 two different types of cosmological dynamics are possible:
 
 \begin{itemize}
  \item 

 In the case where the coupling parameter is small, namely $0<\xi<3/8$,
 only two critical points exist, namely  $Q_3$ and $Q_4$. The  
evolution starts with a massless regime (point $Q_3$ which is unstable node)
and ultimately results to $(0,0)$ point, except for a measure-zero set of initial conditions which ends in 
the saddle point $Q_4$ representing the exponential solution (\ref{aexpQ4}),(\ref{phiexpQ4}). In Fig.~\ref{Fig1}
we depict the phase-space evolution in the $(\alpha,\beta)$ plane, in the case where  $N=2$, $n=2$, $\xi=1/4$, with $V_0=1$, $\kappa^2=1$.
\begin{figure}[hbtp] 
\includegraphics [scale=0.44] {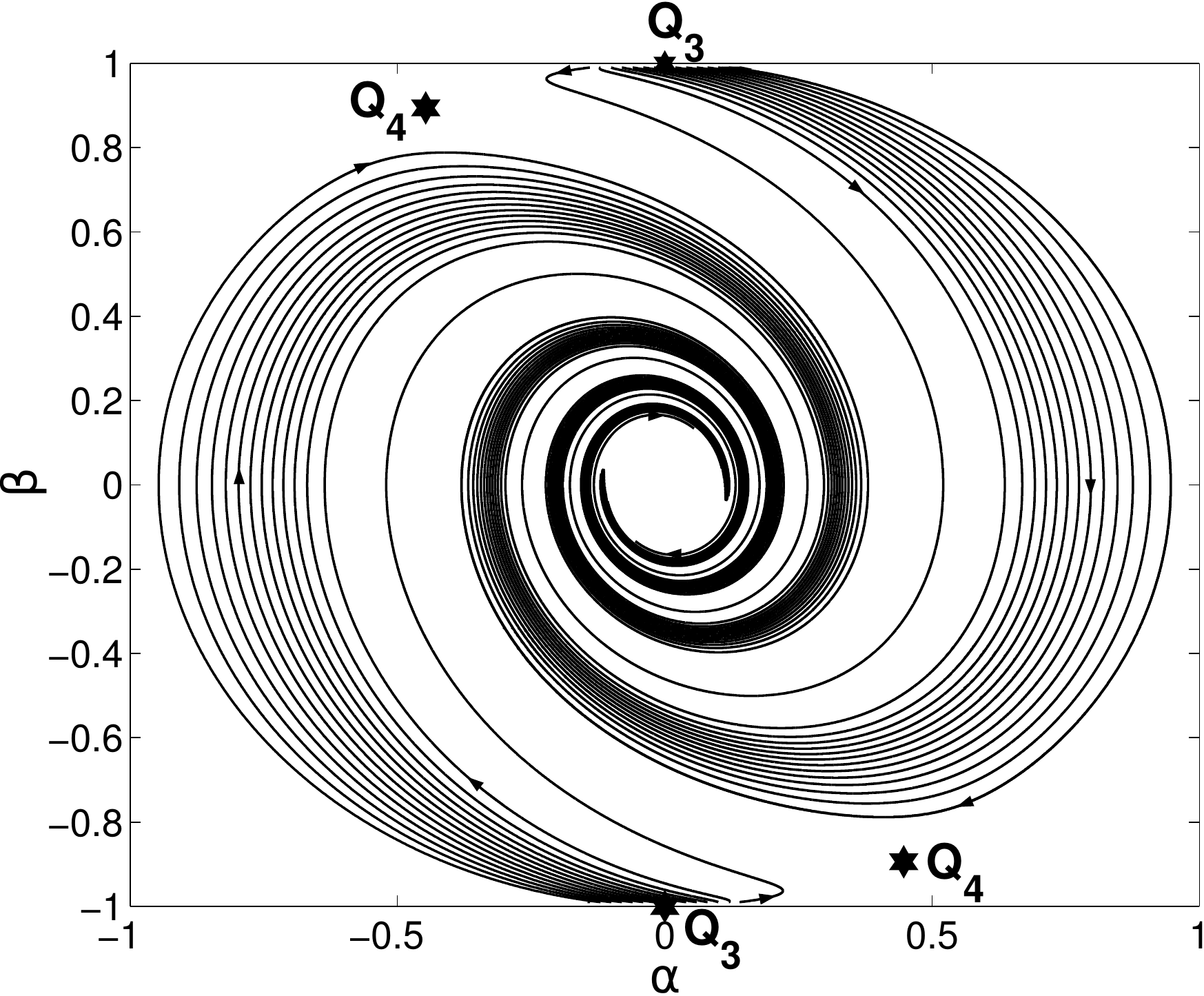}
\caption{
{\it{
The projection of the phase-space evolution of the system (\ref{eq1b}),(\ref{eq2b}) on the  $(\alpha,\beta)$ 
plane, in the case where   $N=2$, $n=2$, $\xi=1/4$, with $V_0=1$,  $\kappa^2=1$, i.e.
in the case of  quadratic nonminimal coupling function   and quadratic potential. }}   }
\label{Fig1}
\end{figure}

  \item 
  In the case where the coupling parameter is large,   namely
   $\xi>3/8$, no critical point exist, and thus the scalar field exhibits
 oscillations. This behavior is depicted in Fig.~\ref{Fig2}, for the case of $N=2$, $n=2$, $\xi=4$, with $V_0=1$, $\kappa^2=1$.
\begin{figure}[hbtp] 
\includegraphics [scale=0.44] {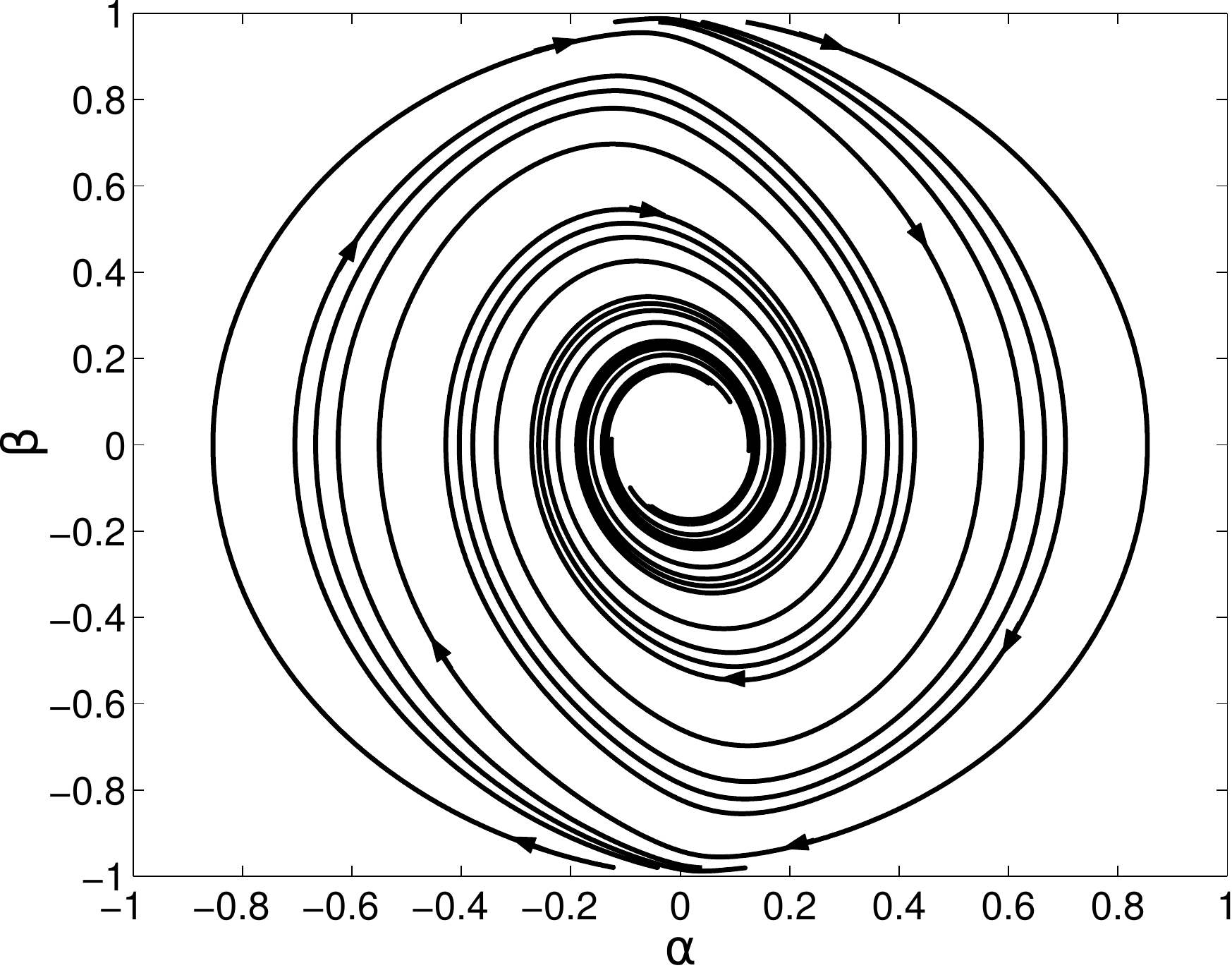}
\caption{{\it{
The projection of the phase-space evolution of the system (\ref{eq1b}),(\ref{eq2b}) on the  $(\alpha,\beta)$ 
plane, in the case where   $N=2$, $n=2$, $\xi=4$, with $V_0=1$,  $\kappa^2=1$, i.e.
in the case of  quadratic nonminimal coupling function   and quadratic potential. There are no stable fixed points, and the scalar field exhibits oscillations. }} 
  }
\label{Fig2}
\end{figure}

 \end{itemize}

Let us now examine the $n>2$ case. In this case, the solutions, when they exist, are power laws instead of exponentials.
Having in mind the existence and the stability conditions of points $Q_1$ to $Q_4$ of subsection \ref{N22}, we deduce that 
 two different types of cosmological dynamics are possible:

  \begin{itemize}

  \item 
  
In the case of small coupling parameter, namely 
$\xi< 6/(n+2)^2$, the phase-space
trajectories begin from the  unstable node $Q_3$ and tend to  dumping oscillations near $(0,0)$ point, since $Q_4$ is saddle.
This behavior can be observed in 
Fig.~\ref{Fig3}, in the case where   $N=2$, $n=4$, $\xi=1/10$, with $V_0=1$, $\kappa^2=1$.
Note that the points $Q_3$ and $Q_4$ coincide, making the resulting phase portrait less
informative.
\begin{figure}[hbtp] 
\includegraphics [scale=0.44] {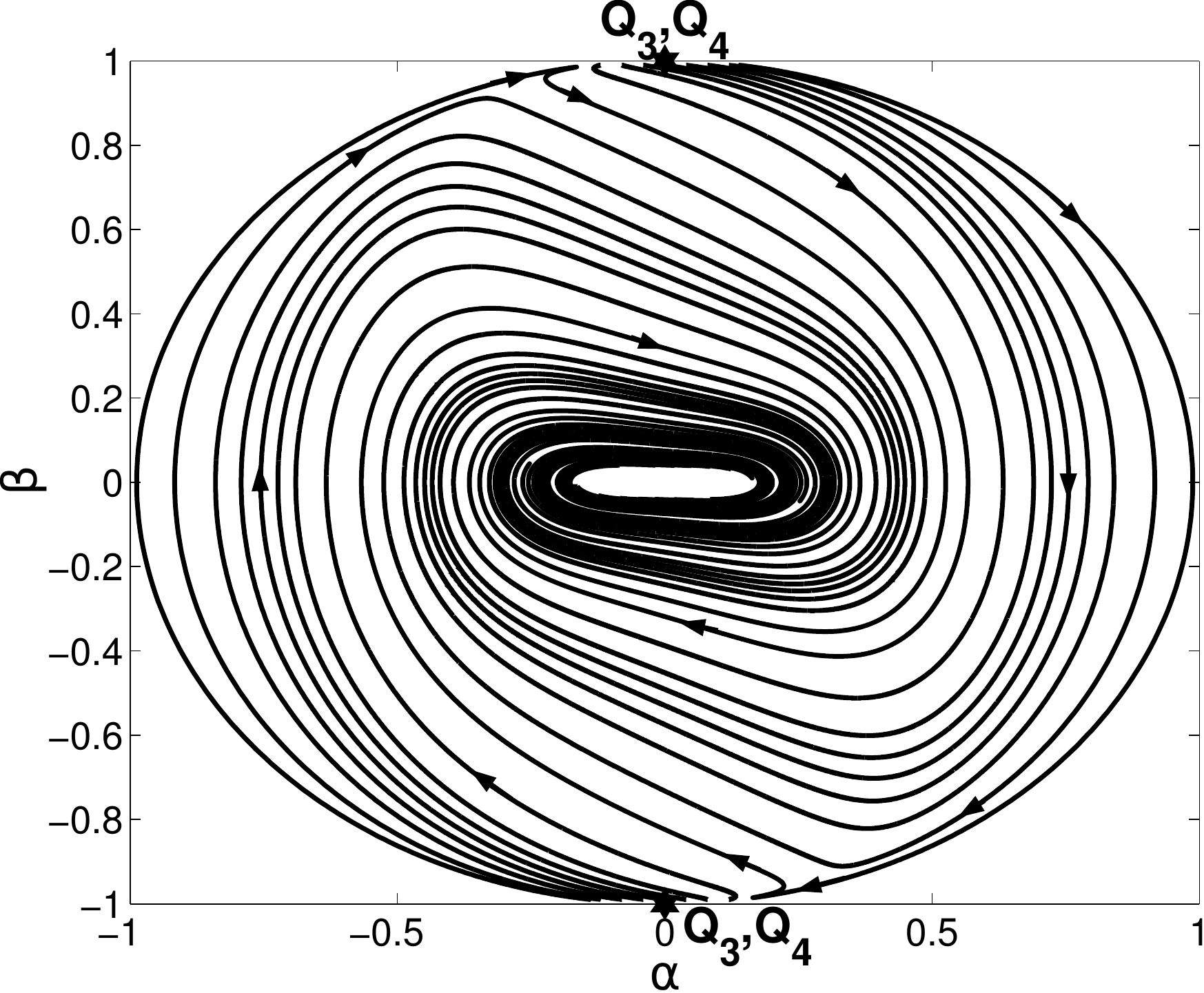}
\caption{
{\it{
The projection of the phase-space evolution of the system (\ref{eq1b}),(\ref{eq2b}) on the  $(\alpha,\beta)$ 
plane, in the case where   $N=2$,  $n=4$, $\xi=1/10$, with $V_0=1$,  $\kappa^2=1$, i.e.
in the case of  quadratic nonminimal coupling function   and quartic potential.}}
  }
\label{Fig3}
\end{figure}

\item

In the case of large coupling parameter, namely $\xi>6/(n+2)^2$,
the critical points do not exist, and thus the scalar field exhibits oscillations. This behavior can be seen
in Fig.~\ref{Fig4}.
\begin{figure}[!] 
\includegraphics [scale=0.44] {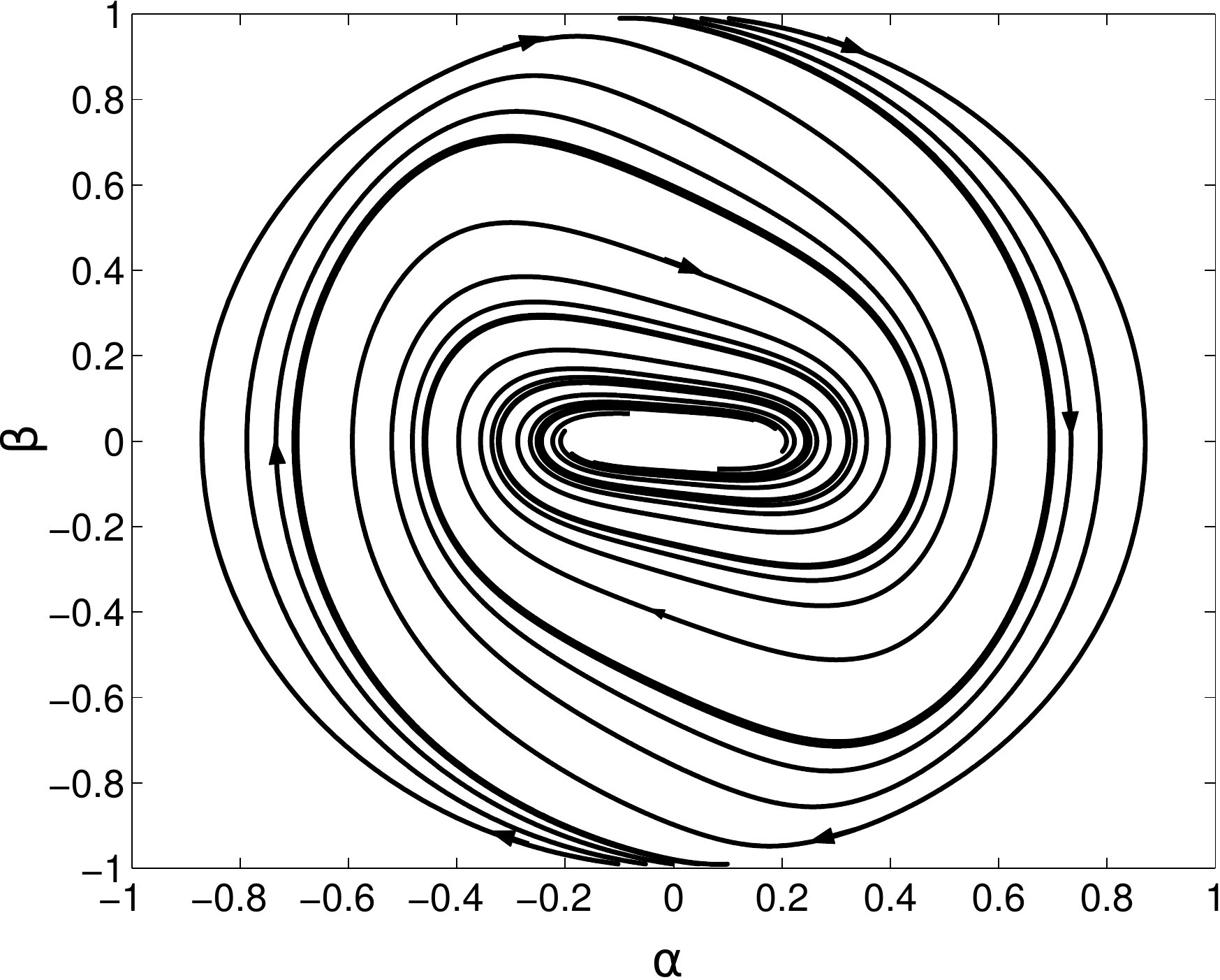}
\caption{
{\it{
The projection of the phase-space evolution of the system (\ref{eq1b}),(\ref{eq2b}) on the  $(\alpha,\beta)$ 
plane, in the case where   $N=2$,  $n=4$, $\xi=1$, with $V_0=1$,  $\kappa^2=1$, i.e.
in the case of  quadratic nonminimal coupling function   and quartic potential.  There are no stable fixed points, and the scalar field exhibits oscillations.}}
 }
\label{Fig4}
\end{figure}
 
\end{itemize}
  
Let us now compare the above behavior with the corresponding one in the scalar-curvature case, which will help to reveal
the special features of the nonminimal coupling in teleparallel gravity. We can easily remark two very important 
qualitative differences concerning possible future and past asymptotic in these two theories
of non-minimally coupled scalar field.  We remind a reader that the action for the scalar-curvature theory is usually written as

\begin{equation}
\label{action}
S=\int d^4 x \sqrt{-g}\left[ \frac12\left(m_{p}^2-\xi B(\varphi)\right)R-
\frac12g^{\mu\nu}\varphi_{,\mu}\varphi_{,\nu}-V(\varphi)\right],
\end{equation}
so it has the same structure as the scalar-torsion theory with curvature instead of torsion.

Firstly, in the case of standard, scalar-curvature models with 
power-law  potentials and coupling functions, the future scalar-field behavior
can be of two possible types: either solutions with $\phi \to \infty$ or damped scalar field oscillations.
The former behavior is realized for
$n<2N$ (in the present paper notations), and moreover it represents a Big Rip singularity if $N<n<2N$ \cite{Sami, Vernov}.
On the contrary, in teleparallel nonminimal coupling with positive even $N$ and $n$
the only stable future asymptotics are the oscillations (note that in this case one needs to be based on the numerical investigation
in order to study the phase-space behavior since 
the scalar field
oscillations cannot be obtained analytically through the dynamical
analysis).
In some sense, the theory under investigation resembles better the theory
with minimally coupled scalar field, where stable future regimes apart from oscillations  are impossible
for simple power-law potentials.

The reason behind this difference is quite clear. In the equation of motion for the scalar field the additional term
originating from the non-minimal coupling, for positive  $\xi$ and even $N$  has the same sign as the potential term for even $n$, and, as a result,
can only enhance the driven force pushing
the scalar field towards the minimum of its potential.
On the contrary,  looking at the equation of motion for the scalar field in the scalar-curvature case

\begin{equation}
\ddot{\phi}=-3H \dot\phi+\frac{3\xi B'(1-3\xi B'')}{1-6\xi B+9{\xi}^2{B'}^2}\dot\phi^2-
\frac{V'+6\xi(2V B' -V' B)}{1-6\xi B+9{\xi}^2{B'}^2},
\end{equation}
 we can see that
additional term in the case of curvature non-minimal coupling has a more complicated
structure, with different signs, and under some conditions can  push the scalar field into infinity.

Hence, it is evident that in order to make non-trivial future asymptotics in the present scenario, we need to balance the  
potential and correction terms and thus we need either to consider
negative $\xi$ or to match an increasing coupling function with a decreasing potential (or vice versa, however this case is more ``exotic'').
For instance, as we saw, for positive even $N$ and $n$ the fixed point $P_1$ corresponding to de Sitter solution exists only for negative $\xi$.
The detailed study of such suitable constructed solutions, based on the interplay between the potential and the coupling
function, lies beyond the scope of the present work. Studies of viability of such scenario for description of present cosmic acceleration 
requires also considering of matter contribution.  We leave these interesting questions for
a future investigation.
       
Secondly, concerning the past behavior, we also see that the  past asymptotics in the scenario at hand
are different from both minimal and nonminimal curvature theory. In both these scalar-curvature theories 
the source point on phase diagrams represents the massless field regime, existing if the potential is not too steep (exponentially 
steep for curvature  minimal coupling \cite {Foster} and $n<5N$ for curvature non-minimal coupling \cite{Sami, Vernov}).
However, in teleparallel gravity with power-law potentials, this regime exists only if  $N$ does not exceed $2$. For  $N>2$ (as well as for $N=2$, 
and $\xi>6/(n+2)^2$) we have infinite scalar
field oscillations near a singularity, similar to those in the theory of a minimally coupled field with exponential potentials. 
Finally, note that in the present scenario we find that the case of $N=n=2$ is exceptional
since only under this condition the unstable exponential solution exists. This feature has no analogies in the curvature theory.

\section{Conclusions}
\label{Conclusions}

In this work we investigated cosmological scenarios in the framework of teleparallel gravity with nonminimal coupling.
Although in the case of absence of the scalar field, or in the case of minimal coupling, teleparallel gravity is completely 
equivalent with general relativity at the level of equations, when the nonminimal coupling is switched on the two theories 
become different, corresponding to distinct classes of gravitational modification. Hence, in order to reveal the differences between 
nonminimal scalar-curvature and nonminimal scalar-torsion theories, we use the
powerful method of dynamical analysis, which allow us to study their global behavior without the need of extracting exact analytical
solutions.

Focusing on the cases of power-law potentials and nonminimal coupling functions, we showed that contrary to the case of scalar-curvature gravity,
teleparallel gravity has no stable future solutions for positive nonminimal coupling, and the scalar field results always in oscillations. Additionally, 
concerning the past behavior, while scalar-curvature exhibits a massless field regime for not too steep potentials, nonminimal scalar-torsion gravity exhibits
this feature for not too steep coupling functions. 

The reason behind these differences is the specific and simple form of the nonminimal correction 
in the scalar field equation in the case of nonminimal scalar-torsion theory,
comparing to the more complicated  corresponding term  in nonminimal scalar-curvature theory. In particular, in the former case the correction term is relatively simple and 
thus it can lead to well-determined changes in the dynamics comparing to the minimal model,  while in the latter case the correction term is complicated and thus it cannot lead to
  well-determined, one-way changes in the dynamics comparing to the minimal model. Hence, since minimal scalar-curvature is equivalent with 
  minimal scalar-torsion gravity, nonminimal scalar-torsion  gravity is relatively close to them, while nonminimal scalar-curvature gravity is radically different.
The significant difference of the two theories was already known  \cite{Geng:2011aj,Xu:2012jf}, however in the present work we verified it analyzing in detail the dynamics of the pure scalar-gravity sector, in order to remove possible effects
of the matter part.

Clearly, the torsional modification of gravity, and its coupling with scalar and matter sectors, brings novel and significant features, with no known curvature
counterparts. Thus, it would be interesting to include these capabilities in the model-building of cosmological scenarios.

\begin{acknowledgments}
The research
of ENS is implemented
within the framework of the Action ``Supporting Postdoctoral Researchers''
of the Operational Program ``Education and Lifelong Learning'' (Actions
Beneficiary: General Secretariat for Research and Technology), and is
co-financed by the European Social Fund (ESF) and the Greek State.
  The work of AT is supported by RFBR grant 14-02-00894, and 
  partially supported by the Russian Government Program of Competitive Growth of Kazan Federal University.
\end{acknowledgments}

\end{document}